# Fabrication, detection and operation of a three-dimensional nanomagnetic conduit

**Dédalo Sanz-Hernández[1], Ruben F. Hamans[2], Jung-Wei Liao[1], Alexander Welbourne[1], Reinoud Lavrijsen[2], Amalio Fernández-Pacheco[1]\*.**

[1] Cavendish Laboratory, University of Cambridge, JJ Thomson Avenue, Cambridge CB3 0HE, UK

[2] Department of Applied Physics, Eindhoven University of Technology, PO Box 513, 5600 MB, Eindhoven, The Netherlands

*af457@cam.ac.uk

## Abstract

**Three-dimensional (3D) nanomagnetic devices such as the vertical racetrack memory have attracted a huge interest due to their potential for memory and logic applications. However, their implementation has not been realised due to great challenges regarding fabrication and characterisation of 3D nanostructures. Here, we show a 3D nanomagnetic system created by 3D nano-printing and physical vapour deposition, which acts a conduit for domain walls. Domains formed at the substrate level are injected into a 3D nanowire, where they are controllably trapped using vectorial magnetic fields. A dark-field magneto-optical method for parallel, independent measurement of different regions in individual 3D nanostructures is also demonstrated. This work opens a new route for the advanced study and exploitation of 3D nanomagnetic systems.**

## Introduction

Magnetic conduit devices based on nanowires (NWs) are one of the most advanced non-volatile nanoelectronic systems ever realised [1–8]. These devices exploit the domain wall (DW) conduit behaviour [10], where DWs flow along NWs as a collection of separated objects, controlled by a wide range of mechanisms (magnetic fields, electrical/spin currents, spin waves, strain…). In these systems, wires are not mere interconnectors, but become the device itself.

Extensive knowledge has been acquired during recent years regarding the physics of DWs in NWs: phase diagrams [9,10], injection, motion and trapping [11,12], logic gates [1,13], new spin-orbit torque mechanisms [4,14,15] and dynamic properties [16,17], are some examples of the great progress achieved. The DW racetrack memory [3] is now a universal concept in the field of spintronics, having motivated, for instance, the recent works on skyrmionic bubble memories [18,19]. However, although two-dimensional (2D) NW devices are now a reality, a 3D NW device such as the vertical racetrack [2] is yet to be realised due to phenomenal patterning and characterisation challenges [20]. Until now, DW motion perpendicular to the substrate plane has been generally studied by means of out-of-plane NWs made by chemical methods and templates [21,22]. This approach is nevertheless restricted to disconnected, vertical straight NWs, and usually involves the release of the wires from the templates for subsequent characterisation. Moreover, higher density of defects and rougher interfaces prevent smooth DW motion in comparison with their planar counterparts [22], where materials grown by physical methods are used. New ways to create and probe high-quality magnetic NWs forming real 3D geometries are therefore essential for further progress in this area.

Here, we show a 3D nanomagnetic conduit created by a combination of direct-write nanolithography and standard physical vapour deposition. DWs are controllably injected from the substrate plane into a 3D NW interconnected at an angle, where they move and become trapped at different locations under the application of external vectorial magnetic fields. A magneto-optical dark-field method which independently probes the substrate and the 3D NW is employed to characterise the nanodevice. The methodology presented here can be extended to complex materials and geometries, paving the way for advanced studies of 3D nanomagnets.

## 3D nanofabrication

The 3D magnetic conduit was created via a two-step lithography process: In step 1, a non-magnetic scaffold is built using 3D nano-printing via Focused Electron Beam Induced Deposition [23] (Fig.1a). In step 2, a magnetic thin film is deposited over the whole sample using thermal evaporation (Fig.1b). In this work, a nano-ramp forming an angle with the substrate is used as the non-magnetic scaffold. Two supporting pillars are added to ensure mechanical stability throughout the whole fabrication and characterisation process. A curved interconnecting section is introduced between the ramp and the substrate to enable a smooth DW transmission from the 2D plane into the 3D NW. Permalloy is chosen as the evaporated magnetic material, due to its good DW conduit properties in 2D NWs [1,2]. A detailed account of the fabrication process and exact geometry of the nanostructures can be found in methods and supplementary information (SI) 1A.

The final nanostructure consists of three parts (Fig. 1c): A magnetic thin film on the plane of the substrate, which acts as a source of magnetic DWs, a 3D NW designed to transport these DWs, and a 2D-3D interconnect region which facilitates their transmission between both. Fig. 1d and 1e show respectively, Scanning Electron Microscope (SEM) micrographs of the main structure studied in this work and of the corresponding interconnect area. Fig. 1f shows an extra 3D structure, in which the NW is disconnected from the film source via shadowing from a nano-bridge fabricated before evaporation. This structure acts as a control during magnetic experiments, as detailed later.

## Magneto-optical detection

Magnetic characterisation of 3D nanostructures is currently under intense investigation [20,24], and presents remarkable challenges. A three dimensional geometry hinders the use of standard magnetometry or magnetic microscopy methods, suitable for planar nanostructures. We recently reported detection of suspended NWs using magneto-optical

Kerr effect (MOKE) [25]. However, the 3D nanostructures studied here do not have any lateral separation between NW and film, which demands a novel approach for detection of both parts independently; this is a requirement to determine their magnetic switching mechanisms. In this work, we have developed dark-field MOKE, exploiting the different angles formed by the two components of the 3D nanostructure (thin film and NW) with respect to the laser beam. Light is collected with two point detectors positioned at two different angles, those result of specular reflections from either of the two areas. With this optical arrangement, we detect the magnetic switching of a single 3D NW and the film around it, independently and simultaneously (see Fig. 2a). This overcomes detection limits of a standard MOKE configuration, where Kerr signal is completely dominated by the thin film (SI 1B and 1C). Magneto-optical detection at different angles has been employed previously in transmission configuration for films with varying refractive index [26], and in diffraction-MOKE of nano-element arrays [27]; here, individual 3D nanostructures are detected by using their geometry to distinguish parts which are not separated laterally, but lie at different angles to the incident laser. The experimental configuration includes three sets of coils, making it possible to perform complex magnetic field sequences in 3D. A complete description of the optical setup can be found in SI 1D.

The successful implementation of dark-field MOKE is illustrated in Fig. 2b-e. Fig. 2b and 2c show optical images of the sample collected with a CCD camera positioned at the angle corresponding to the black detector shown in Fig. 2a (i.e. only laser reflections at that angle – corresponding to the NW – are observed). In Fig. 2b the laser spot is positioned on top of a 3D NW (black dots), resulting in an image with a well-defined laser reflection. This is contrary to Fig. 2c, where the laser beam is placed 20 µm away from the NW (dashed circle), leading to no laser reflection. Additionally, Figs. 2d and 2e show hysteresis loops measured by the blue and black detectors upon application of magnetic fields. The magnetic switching of NW and thin film occur at very distinct field values, with no observable cross-contamination between the two measurements.

## Domain wall conduit behaviour in 3D nanowires

The switching mechanisms of 3D NWs under external magnetic fields have been investigated using dark-field MOKE. The complex geometry of the nanostructures makes it challenging to interpret isolated magnetic hysteresis loops, since any applied field has a different projection along different parts of the nanostructure. In order to aid interpretation, identical experiments have been carried out for the disconnected and interconnected nanostructures of Fig. 1f and 1d. Fig. 3 summarises this study, including the types of magnetic fields employed (Fig. 3a), as well as hysteresis loops (Fig. 3b and 3c) and switching field diagrams (Fig. 3d and 3e) for the two nanostructures, respectively. Additionally, Fig. 3f-i show results from micromagnetic simulations in an interconnected 3D system such as the one investigated experimentally, for comparison with Fig. 3e.

Magnetic fields consist of an oscillating x-field ($H_x$) along the NW axis, combined with constant z-offsets ($H_z$) normal to it (along its thickness direction) (Fig. 3a). Such fields have the same positive and negative projections along any element that lies in the xy plane. Conversely, those projections differ for any element not lying in that plane. This is observed for the disconnected structure in Fig. 3b: NW MOKE loops are symmetric as a function of $H_x$ for any $H_z$, but become shifted (geometrically biased) for the film, when a finite $H_z$ is applied. The differing symmetry of both components, including the geometrical bias of the film, is readily observed in a $H_x$-$H_z$ switching diagram (Fig. 3d). Here, red and turquoise dots correspond to the switching field values for NW and film, respectively. For such a disconnected system, both NW and film switch via the nucleation of domains, independently from each other. The resulting switching lines are therefore perpendicular to the plane of each component: Vertical lines for the NW ($\mu_0 H_N \simeq 5.5$mT for all $H_z$), and slanted (very close to each other – 0.2 mT for $H_z = 0$) lines for the film (see the inset of Fig. 3d for axes and normal directions of both). As expected, the film switching occurs at much lower values than for the NW, due to the magnetic softness of Permalloy and the lack of lateral confinement in the film. The negligible $H_z$-dependence observed for NW nucleation lines is also consequence of the small $H_z$ field values applied in comparison to its demagnetising field along the z-direction, and is consistent with nucleation via curling of a small oblate ellipsoid (See Methods).

In the case of the interconnected nanostructure (Fig. 3c and 3e), the magnetic behaviour of the film is identical to previously, whereas the NW now presents a richer switching landscape. This includes asymmetric MOKE loops (Fig. 3c), and a switching diagram formed by four main types of field events (Fig. 3e). For large $H_z$ offsets, the situation is analogous to previous studies in 2D L-shaped NWs under equivalent field routines [28], presenting very asymmetric hysteresis loops (See $\mu_0 H_z$ = +11.3mT in Fig. 3c). This is a consequence of the large geometrical bias in the film, which prevents it from switching under $H_x$ before the NW. The NW thus reverses back and forth via either the propagation of a DW located between film and NW, or via the nucleation of new domains due to the absence of DWs [28]. This depends on the relative sign of $H_x$ and $H_z$. In this NW, switching via nucleation is substantially sharper than via propagation. Since both types of events occur within the NW plane, straight lines with equal slope are observed (red and blue dots in Fig. 3e). However, they are not perfectly perpendicular to the x-direction, as for nucleation in the disconnected NW (Fig. 3d), but have a lower slope. We associate this deviation to geometrical waviness in the connected NW, as observed by SEM (SI 1A). The range of possible axes for the NW inferred from SEM measurements are indicated in the inset of Fig. 3d, with both nucleation and propagation lines falling within this angular range. Taking this into account, nucleation lines fit well to the curling model [29], with same nucleation volume as the disconnected NW (Methods). Moreover, we can define values for nucleation and propagation fields ($H_N$ and $H_P$, respectively), by measuring the distance from either switching line to a parallel which crosses the origin, as indicated in the diagram. We obtain $\mu_0 H_N$ = 8.4 mT and $\mu_0 H_P$ = 2.8 mT.

For smaller $H_z$, a different regime is accessed, since the film now switches before the NW and acts as a source of DWs for both positive and negative $H_x$. A remarkable asymmetry is still observed in the hysteresis loops for this regime (see $\mu_0 H_z$ = +5.6 mT in Fig. 3c). Green and blue dots in the diagram of Fig. 3e represent the two (positive and negative) switching fields. From the different slopes and sharpness, we infer that the green switching is in fact the continuation of propagation events, now for smaller $H_z$ offsets. The symmetry of the blue lines indicates, however, switching within a different plane, located in-between film and NW. Following geometrical arguments, this corresponds to the interconnect between both (Fig. 1d), as detailed later. We thus associate a transmission field ($H_T$) to this type of switching, which characterises DW depinning at the interconnect area, with $H_T$ = 2.0 mT defined analogously to $H_N$ and $H_P$ (see Fig. 3e). The dependence observed for both propagation and transmission lines fits well with the Kondorsky model [30] for DW motion (Methods). Remarkably, with $H_N$ > $H_P$, $H_T$, the injection of DWs from the film into the 3D NW is realised, with the system operating as an effective 3D conduit of DWs.

In order to understand in detail the mechanisms for DW motion after injection from the film, we have performed micromagnetic simulations [31] (Methods). For this, we have investigated the depinning of vortex DWs at the interconnect area (Fig. 3,f-i) by applying $H_x$-$H_z$ fields from zero to 15 mT at different angles. The 3D geometry numerically investigated mimics the key features observed in the experimental nanostructures, as observed by SEM. This includes a small narrowing in the NW geometry at two locations: one at the start of the interconnect, and a second at an intermediate location within the NW, 600 nm above its base. These two defects are expected to act as DW pinning areas [11]. Fig. 3i shows a diagram generated from the simulations, where each point corresponds to one simulation step. Three different colours denote the different magnetic states observed, as shown in Fig. 3f-h. DW pinning at the interconnect area (Fig. 3f) and at the intermediate defect (Fig. 3g) are represented as blue and green open circles, respectively, and a completely reversed NW (Fig. 3h) as grey open circles. The values and symmetry obtained for transmission and propagation lines in Fig. 3i (green and blue lines) are in good agreement with experiments. Simulations show how the relative sign of $H_x$ and $H_z$, and therefore the differing projections of the total field over the NW and interconnect, determines whether the NW switching is dominated by either transmission or propagation. For positive $H_z$, the NW switches in a single step (from Fig. 3f to 3h): the DW is transmitted from the interconnect and is not pinned at the intermediate defect, due to the large projection of the field over the NW. This is contrary to negative $H_z$ cases, with a small field projection. Here, the DW is first depinned from the interconnect and gets trapped at the intermediate defect (Fig. 3g), until DW propagation takes place, leading to full switching (Fig. 3h). The video of SI 2 illustrates the process in more detail.

In addition to the geometrical depinning effect [32] reproduced by simulations, experiments (Fig. 3e) show a slightly different slope for the two transmission branches. This creates a significant asymmetry in the diagram at the areas marked * and **: Whereas * denotes a direct crossover between transmission and nucleation, ** switching fields deviate (orange points) from transmission, to follow instead the switching line of the film, before the crossover with nucleation. This is a direct consequence of the film acting as a source of domains for the NW at this region of the diagram – film-limited transition. The different slopes for the transmission lines can be understood in terms of DW chirality-dependent depinning [27,33], with $H_z$-chirality selection due to the interconnect curvature, as previously reported for Permalloy 3D curved elements [34].

8/21

### 3D conduit operation

The above results demonstrate a good DW conduit behaviour for the interconnected 3D NWs under investigation. They also show how DWs injected from the 2D film get trapped at different areas of the conduit, depending on the magnitude and direction of the applied field with respect to the three elements of the system (film, interconnect and NW). This has been exploited for advanced field-driven operation of the nanodevice, as shown in Fig. 4a. The figure shows an experiment where $H_x$-$H_z$ rotating magnetic fields (red and blue lines) have been applied, overlaid onto the diagram of Fig. 3e. These fields result in symmetric hysteresis loops, with the NW switching either sharply via DW transmission (Fig. 4b) if the field is counter-clockwise (CCW), or progressively via propagation (Fig 4c) if it is clockwise (CW). The sense of rotation of the field determines the order in which transmission and propagation lines are crossed, allowing us to control if the DW injected from the film gets trapped, or not, at defects along the NW. This property could be exploited to select different functionalities of the 3D nanomagnetic conduit [20]: Acting either as a magnetic 3D interconnector when no trapping is produced within the NW, linking planes hosting functional DW devices; or as a 3D field-driven racetrack memory component, with NW width modulation utilised to trap DWs along its length.

Finally, we have employed fully three dimensional magnetic fields, by combining a rotating $H_x$-$H_z$ with $H_y$ offsets (Fig. 4d). As before, under moderate $H_x$-$H_z$ CCW rotating magnetic fields and $H_y = 0$, the NW switches back and forth via the injection of domains from the film (Fig. 4f), which get transmitted through the interconnect. This results in an alternating NW MOKE signal as a function of time (Fig. 4e). On the contrary, the addition of $H_y$ magnetically saturates the film (Fig. 4g), inhibiting the injection of DWs. As the applied rotating field is not enough to nucleate domains in the NW, no switching is produced, and a constant low MOKE signal is obtained in this case (Fig. 4e). This result encapsulates the full vectorial nature of the 3D system, where the application of a $H_y$ offset acts as a magnetic gate, controlling the injection of DWs from the 2D film into the 3D NW, and exemplifies how the full xyz magnetic energy landscape can be exploited in 3D nanomagnetic systems.

**Conclusion**

We have demonstrated a new platform for the fabrication and study of three dimensional magnetic nanostructures. High quality magnetic material has been successfully shaped in three dimensions by thermal evaporation onto a non-magnetic scaffold fabricated using 3D nano-printing. Different areas of the system can be detected independently using a dark-field magneto-optical method tailored to the 3D geometry of the nanostructure. A nanomagnetic conduit which can transport magnetic information in three dimensions has been demonstrated, and the mechanisms driving such a process characterised. The methods demonstrated in this work can potentially be extended to complex 3D geometries and nanostructured multi-layered materials. This development sets the path for new spintronic devices that can take advantage of the three dimensions, increasing functionality and density. Novel possibilities also open up in fields of study such as magnetic spin-ice and magnetic neural networks, in which a much greater degree of connectivity between elements would be allowed. Furthermore, the applicability of the techniques demonstrated is extensible to any nanotechnology area which may benefit from extending into the third dimension.

## Acknowledgements

This research is funded by an EPSRC Early Career Fellowship EP/M008517/1, a Winton Fellowship and by a European Erasmus Mobility program. DSH acknowledges funding from a Girton College Pfeiffer Scholarship. We thank Máximo Sanz-Hernández for fruitful discussions.

## Author Contributions

DSH and AFP designed the experiments. RFH, DSH and AFP fabricated the samples; JWL assisted in the fabrication process; DSH and RFH performed the experiments; DSH and AFP analysed the data and wrote the manuscript. All authors discussed the results and contributed to the writing of the manuscript.

# Methods

## Fabrication

The non-magnetic scaffold was fabricated in a SEM/FIB microscope using Trimethyl (methylcyclopentadienyl) platinum as a gas precursor. Growth conditions: 30 kV acceleration voltage, 25 pA nominal beam current, base pressure of $2\times10^{-6}$ mbar and growth pressure of $7\times10^{-6}$ mbar. Stream files specifying substrate coordinates and dwell time were employed at a magnification of 35.000x. The growth is performed with a substrate tilt of 30° in the direction of the ramps. The angle formed by the ramp with the substrate is controlled by scanning the electron beam at a constant speed $v_s$ [35]. Slope, length and width of the NWs studied here were set to 30°, 5 µm and 300 nm respectively, values designed for compatibility with magneto-optical detection.

First, two support pillars of 300 nm x 60 nm cross section and heights of 3 µm and 2.2 µm are deposited by scanning the beam in a single pixel line with a 20 nm pitch. Then the ramp is grown by scanning the beam in a serpentine pattern with longitudinal pitch of 1 nm and transverse pitch of 20 nm. The dwell time is increased from 0 ms to 20 ms over the first 500 nm of patterning, keeping it constant for the rest of the deposit. This generates a high-quality substrate-to-scaffold transition, which is transferred to the magnetic thin film upon thermal evaporation. The shadowing nano-bridge of the control structure was grown as the intersection of two shorter ramps using the same conditions.

A 50 nm Permalloy thin film was thermally evaporated perpendicularly to the 2D substrate, at a rate of 1.7 nm/min and growth pressure of 4.8 x $10^{-6}$ mbar. Chamber base pressure: 2.5 x $10^{-7}$ mbar. After evaporation, NWs showed small roughness, with slope deviations from their mean value of up to 10˚.

**Magneto-Optical Characterisation**

We employ a p-polarised laser beam, which is incident at 45˚ to the plane of the NW (xy plane) and 12˚ to the plane of the substrate. The setup is aligned so that NW, laser and normal to the substrate lie in the xz plane, resulting in two reflected beams, also in the xz plane. These two beams correspond to the reflections from the NW and the substrate, and are picked up by the black and blue detectors respectively (see Fig. 2a). A magnetic quadrupole located in the xy plane is used to apply $H_x$ and $H_y$ fields of up to 30 mT and a single pole coil can be used to apply $H_z$ fields of up to 15 mT. Three lenses (not shown) are used to focus the laser spot on the substrate (full width at half maximum ≈ 5 μm) and collect the two reflected beams. Changes in MOKE ellipticity are measured using a combination of λ/4 plate and analyser for each reflected beam. The biggest sources of uncertainty in the measurements are the angle between the substrate and the xy plane, which was set to 33˚±3˚ to better match NW angles obtained experimentally (See SI 1A), and the distance between substrate and z coil, which introduces a 10% uncertainty for $H_z$. Nanostructure and film are independently detected with SNR of 3.6 and 12 respectively. Switching events in the diagrams are defined as a sign change of MOKE signal. SI 1D shows a schematic of the complete optical setup, greyed out components are not used. Lamp is turned on only during alignment.

**Fit of nucleation fields to the curling model**

Domain wall nucleation within magnetic NWs was fitted to the curling model for an oblate ellipsoid with minor and major semiaxes a and b respectively, in which the nucleation field is given by [29]:

$$\mu_0 H_N(a,b,\Omega) = \frac{\mu_0 M_S \left(2D_a(a,b) - \frac{k}{S^2}\right)\left(2D_b(a,b) - \frac{k}{S^2}\right)}{2\sqrt{\left(2D_a(a,b) - \frac{k}{(S(a))^2}\right)^2 \sin^2(\Omega) + \left(2D_b(a,b) - \frac{k}{(S(a))^2}\right)^2 \cos^2(\Omega)}}$$

Where Ω is the angle between the applied field and the minor semiaxis of the ellipsoid, $M_s$ [A/m] is the saturation magnetization, $D_a(a,b)$ and $D_b(a,b)$ are the demagnetizing factors of the ellipsoid along its minor and major axes respectively, k is a geometrical parameter between 1.3793 and 1.424 [36], and S = a/$R_0$ with

$$R_0 = \sqrt{\frac{2\pi A_{ex}}{\mu_0 M_S^2}}$$

$A_{ex}$ [J/m] being the exchange stiffness. Defining m=a/b the demagnetizing factors $D_a(a,b)$ and $D_b(a,b)$ are given by [37]

$$D_a = \frac{1}{1-m^2}\left(1 - \frac{m\cos^{-1}(m)}{\sqrt{1-m^2}}\right); \quad D_b = \frac{1}{2}(1-D_a)$$

We fit the nucleation volume assuming an oblate ellipsoid which is exactly contained within the thickness of the NW, i.e. 2a = cos(30º) × 50 nm, where 50 nm is the thickness of Permalloy deposited perpendicularly to the 2D film, and 30º the nominal angle between the NW and the 2D film. $M_S$ = 8 x $10^5$ A/m and $A_{ex}$ = 1.3 x $10^{-11}$ J/m. As described in the main text, nucleation fields at Ω = 90º were measured as the distance between nucleation lines in Figures 3d and 3e and a parallel passing through the origin. $\mu_0 H_N$ = (5.7 ± 0.4) mT and $\mu_0 H_N$ = (8.4 ± 0.1) mT, for disconnected and connected NWs, obtaining estimations for b of 80 nm and 78 nm respectively.

**Fit of transmission and propagation fields to the Kondorsky model**

The angular dependence of Kondorsky-type domain wall propagation is given by [37]

$$H_{sw}(\theta) = \frac{H_0}{\cos(\theta)}$$

where θ is the angle of the applied field. This relationship is equivalent to straight lines in an $H_z$-$H_x$ diagram. For both propagation and transmission lines, $H_0$ values were determined by measuring the distance between each relevant line and a parallel passing through the origin, as explained in the main text.

## Micromagnetic simulations

Micromagnetic simulations were performed using version 3.9 of the software Mumax3 [31] on an Nvidia GTX1070 graphics card. A cell size of 5 nm was chosen and a total volume of 512 x 256 x 128 cells (2.56 µm x 1.280 µm x 0.64 µm) was simulated. The NW slope was set to 40º with respect to the 2D film. The thickness of the structure is 50 nm perpendicularly to the 2D film to emulate the Permalloy evaporation process. Exchange stiffness $A_{ex}$ = 13 x $10^{-12}$ J/m and saturation magnetization $M_{sat}$ = 8 x $10^5$ A/m. Two constrictions decreasing the width of the ramp by 5 nm at each side (1 cell) were included in the ramp and the interconnect (SI 1E), as determined by SEM micrographs. When reaching the 2D film, the lower constriction penetrates 20 nm into it.

The system was initialised with a random circular patch 250 nm in radius at the base of the interconnect. The magnetization was then relaxed using the built in relax() function of Mumax3 obtaining a Vortex domain wall (Fig. 3f). The stray fields from a larger 2D film and ramp were imported at the boundaries of the system to avoid the formation of closure domains. The relaxed configuration was stored and used to initialise all field sequences.

Field sequences from 0 to 15 mT were applied at different angles in 40 steps, performing a steepest gradient minimisation after each increase using the built-in function minimize() in Mumax3. The minimiser_stop parameter was set to $10^{-6}$.

The maximum angle between spins remained below 43˚ throughout the whole simulation, corresponding to a maximum error of ≃4% in exchange energy [31]. The video in SI 2 shows the evolution for two field sequences at positive and negative $H_z$ offsets. The complete script used to perform the simulations is available via the SI.

# Figures

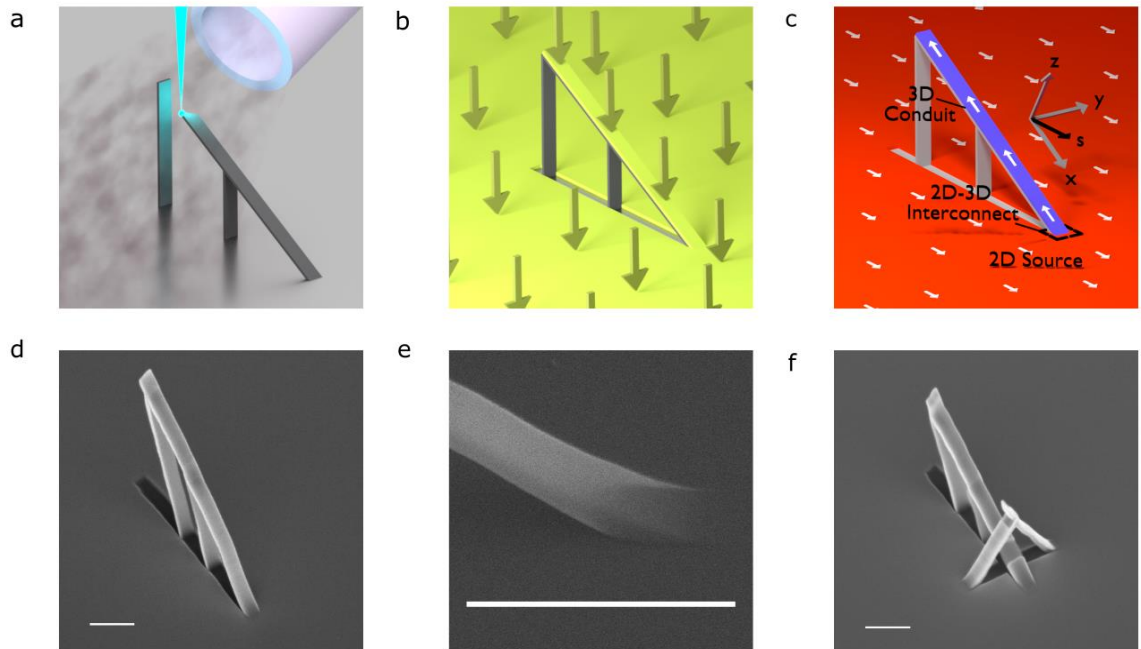

**Figure 1 ¦ Fabrication of a 3D magnetic domain wall conduit. a**, 3D printing of a non-magnetic scaffold using Focused Electron Beam Induced Deposition. **b**, 3D magnetic nanowire created by depositing a magnetic layer using thermal evaporation. **c,** Schematic of the nano-magnetic system designed. The magnetic state of the 2D source (red) can be transferred into the 3D conduit (blue) via a 2D-to-3D interconnect. **d,** SEM image of a nanomagnetic conduit after Permalloy evaporation (50 nm). The shadow created by the scaffold after evaporation is observed **d**, SEM image of the substrate-to-scaffold connection. **e**, SEM image of a control nanostructure, disconnected from the 2D source by adding a nano-bridge at its base, which shadows the growth of evaporated magnetic material on the area beneath it. Scale bars 1μm.

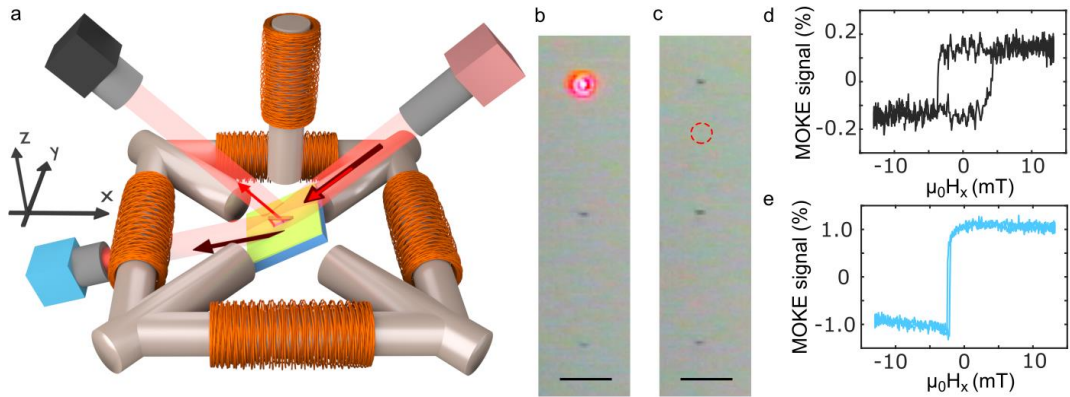

**Figure 2 ¦ Magneto-Optical Detection. a,** Dark-Field Magneto-Optical Kerr effect setup to detect independently a magnetic thin film and 3D NW under vectorial external magnetic fields. A single laser is directed at 45˚ to the xy plane, defined by the nano-ramp. Most of the signal is reflected from the substrate into the blue detector (bright field). The contribution from the NW is detected by the black detector with no contribution from the film (dark field). The angle between the substrate and xy planes is 33˚±3˚, matching the NW inclination. **b,** Optical microscopy image obtained with a CCD camera positioned at the same angle as the black detector. White light is sent to the sample at the appropriate angle. Three NWs are observed as black dots. The laser light is observed as reflected from a NW. **c,** The sample is moved 20 µm up, resulting in no laser light observed by the camera (laser position marked with dotted circle). **d,e,** NW (d) and film (e) hysteresis loops measured in parallel using this setup, upon application of a 14 mT oscillating x-field and a constant z-field offset of 3.2 mT. Scale bars 20 µm.

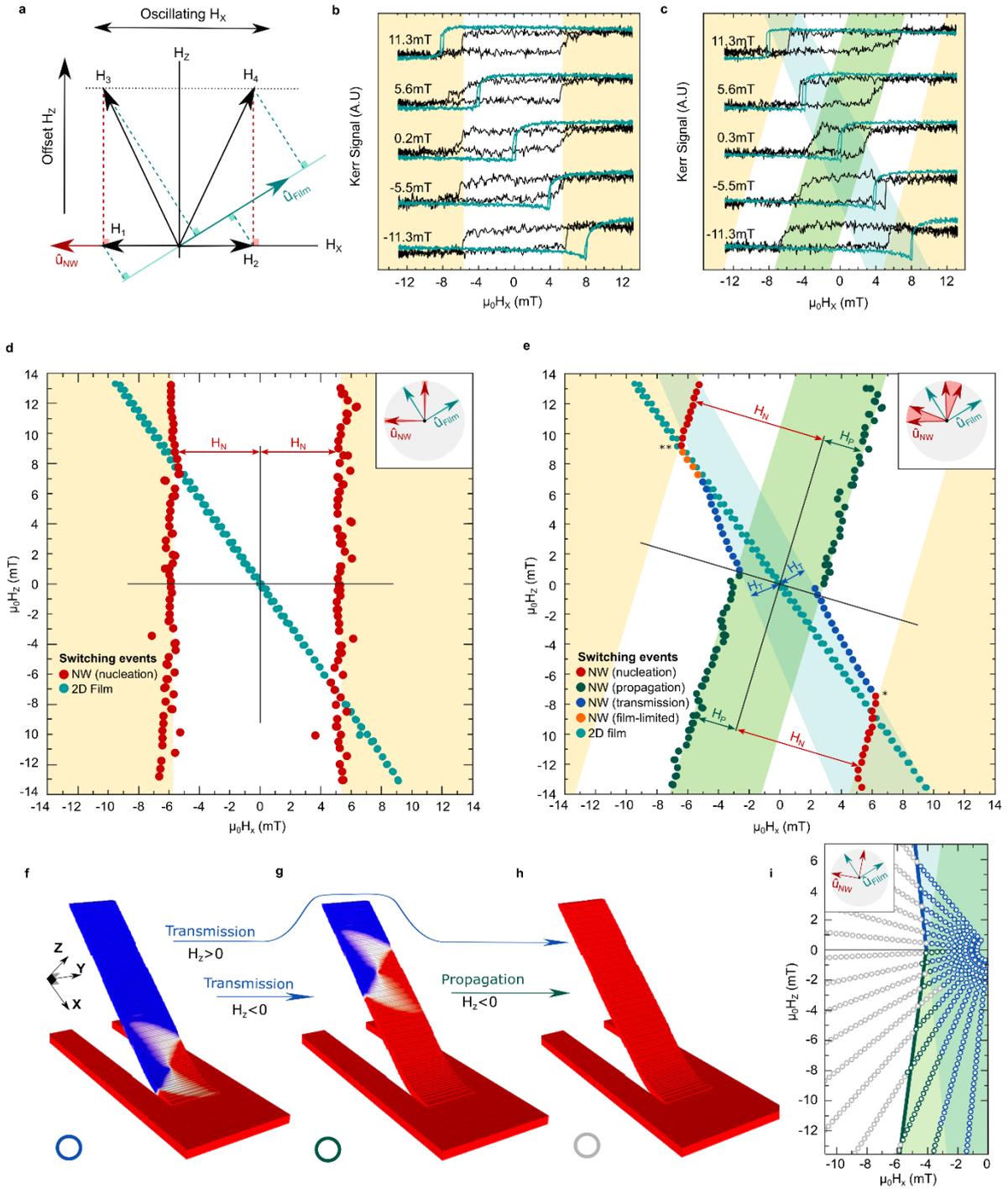

**Figure 3 | Magnetic Characterisation. a,** Effect of an offset field ($H_z$) on the projections of an oscillatory field ($H_x$) onto the NW and film. All projections are symmetric about $H_x = 0$ except for $H_3$ and $H_4$ onto the film. **b, c** Subset of hysteresis loops at different $H_z$ offsets, for disconnected (b) and connected (c) NWs. Turquoise (black) lines corresponds to 2D film (3D NW) signals detected simultaneously using dark-field MOKE. Shading, as in 3c,d diagrams, guides the eye. **d,** ($H_x$, $H_z$) diagram of switching fields for a disconnected NW and underlying 2D film. Switching fields align perpendicularly to their corresponding structure. Areas where the NW has (not) switched are indicated with light orange (white). **e,** ($H_x$, $H_z$) diagram of switching fields for a NW interconnected to a 2D film. NW points are colour separated depending on the switching process. The range of fields for which transmission (propagation) is not possible are shadowed light blue (green). Fields for which nucleation is possible are shadowed light orange. **f-i,** Micromagnetic simulations investigating the depinning of vortex walls from the interconnect area under fields applied at different angles. Snapshots of the simulations include domain wall pinned at interconnect (f), at the NW (g) and the fully-switched system (h). The two possible NW switching processes are indicated for $H_z > 0$ (single step: transmission) and $H_z < 0$ (two steps: transmission + propagation). **i,** Switching phase diagram obtained from simulations, to be compared with left side of Fig. 3e. Point colours match states shown in 3f (blue), 3g (green), 3h (grey). Green and blue lines correspond to switching via propagation (two-steps) and transmission (one step), respectively. Colour shading as in 3e to guide the eye. **Insets (d,e,i):** Directions for nanowire and film axes, and their normal vectors, as determined from SEM images. Angular range where nucleation and propagation can take place is indicated by light red shading.

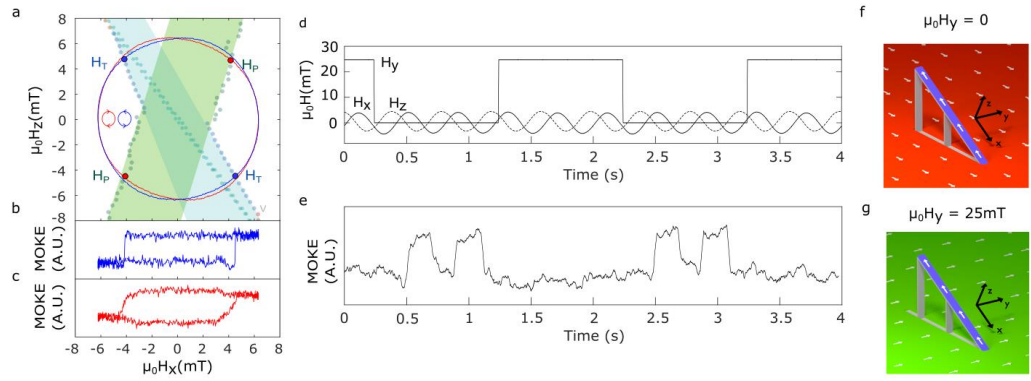

**Figure 4¦ System Operation. a,b,c** Control over switching by application of rotating magnetic fields. Fields applied are plotted as red and blue lines, with corresponding MOKE signals below (b,c). Points where switching occurs for each field sequence are marked by red and blue dots correspondingly. The diagram of Fig. 3d is superimposed for reference. **d,** Magnetic fields applied as a function of time. A transverse field ($H_y$) is employed as a magnetic gate to control the injection of domains from the film, which are transmitted using rotating ($H_x$, $H_z$) magnetic fields. **e,** Corresponding time-dependent MOKE signal. **f,g,** Schematic of the magnetic configuration of the system before NW switching without (f) and with $H_y$ (g).

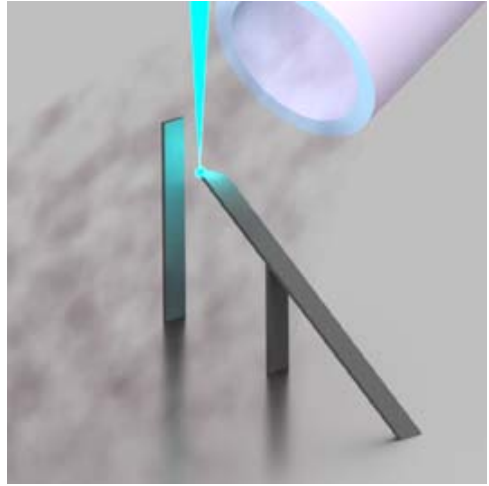 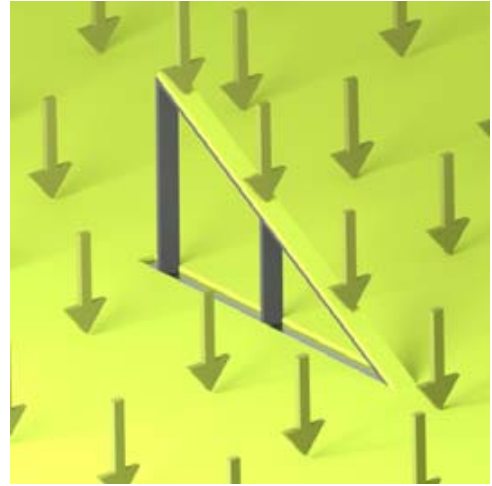 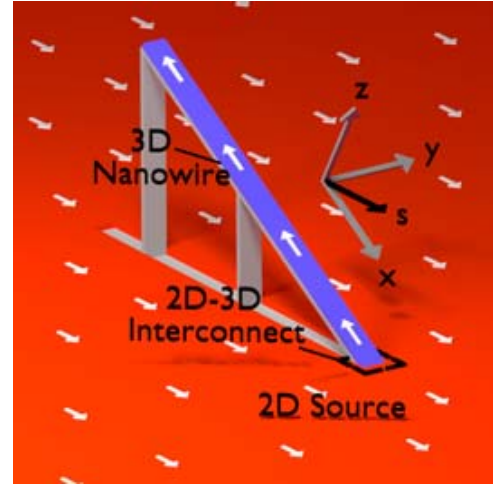
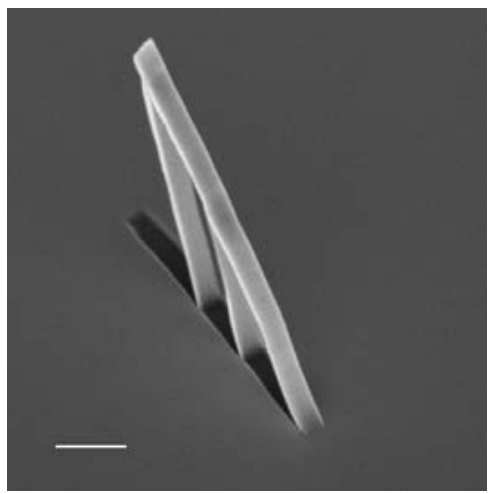 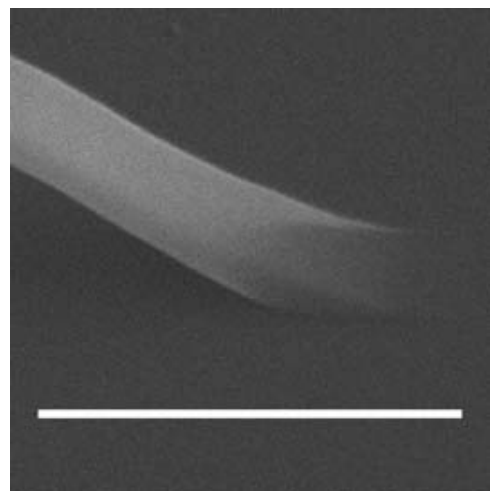 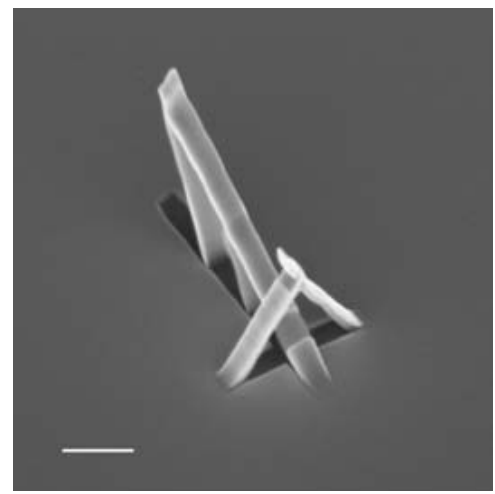

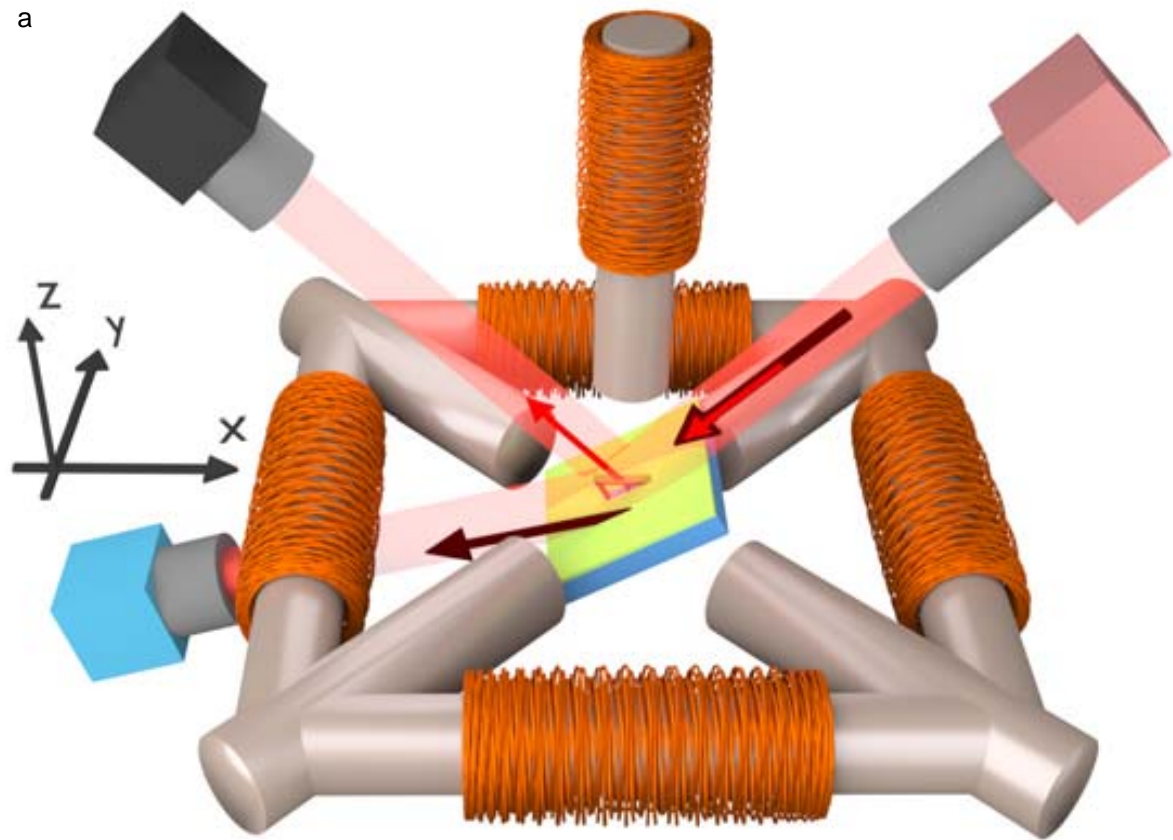 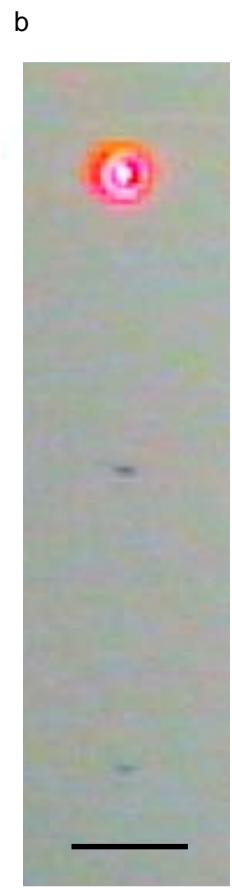 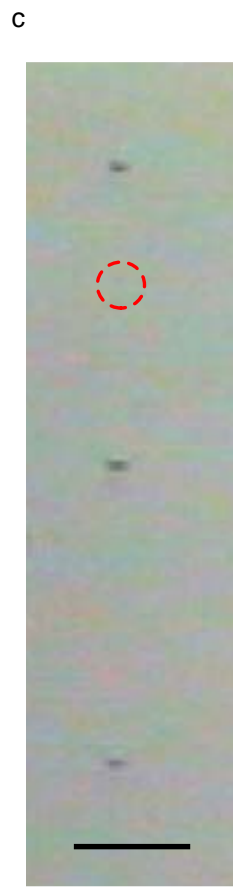 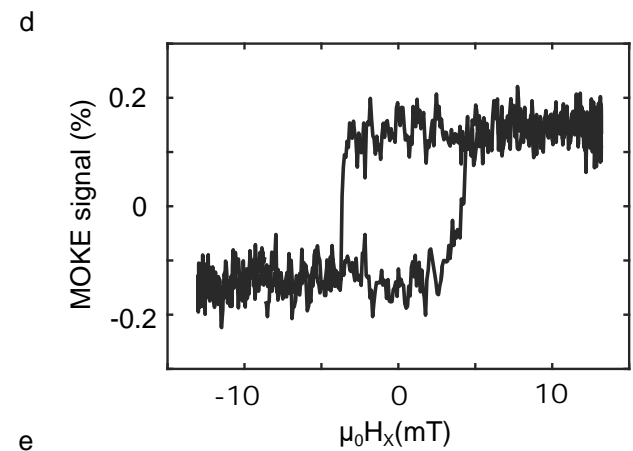 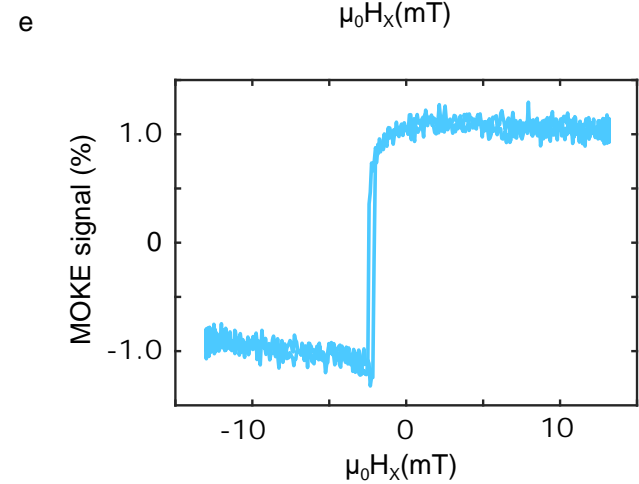

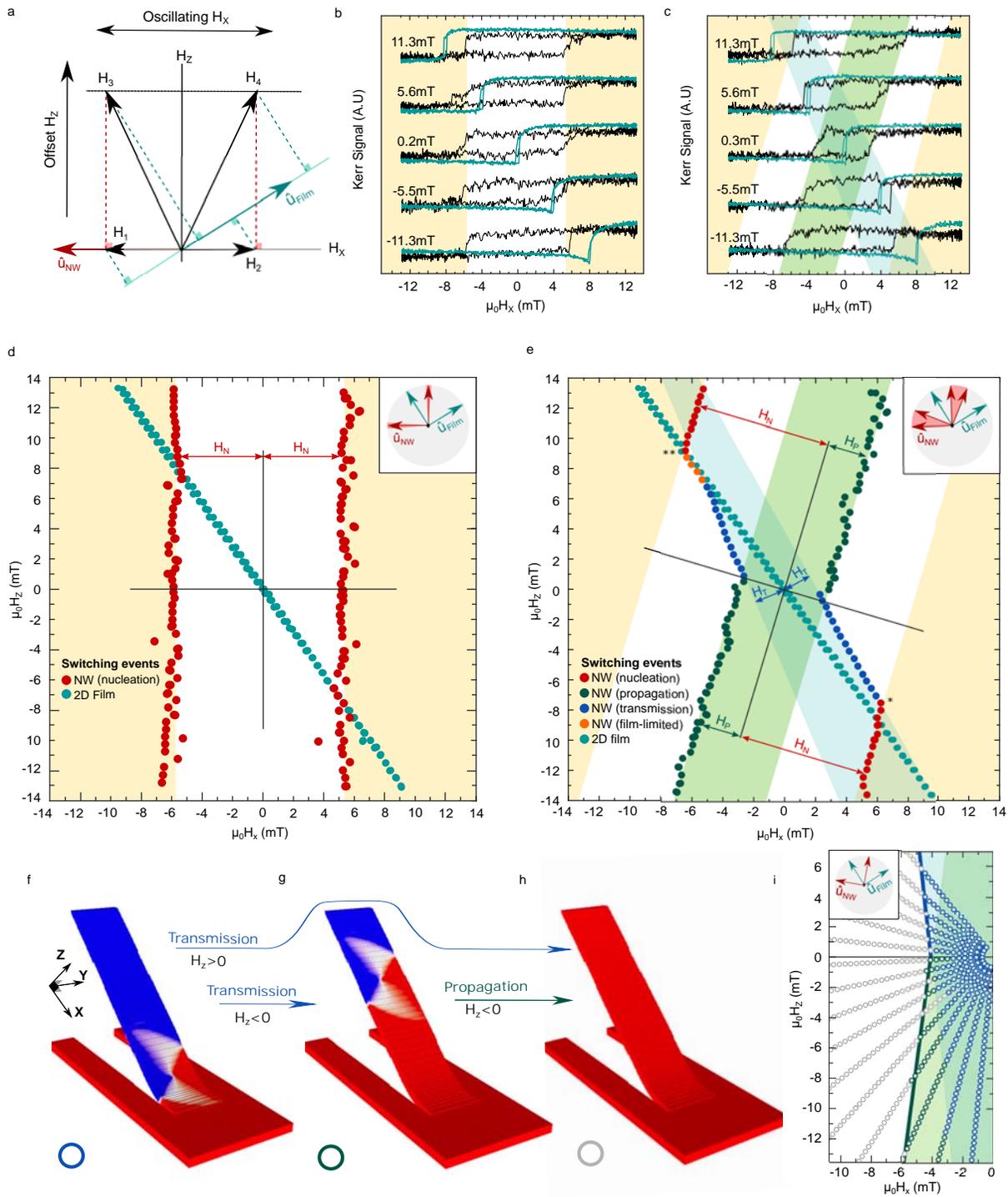

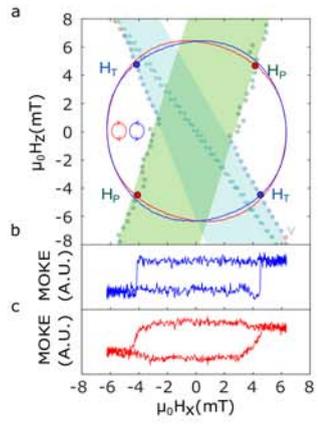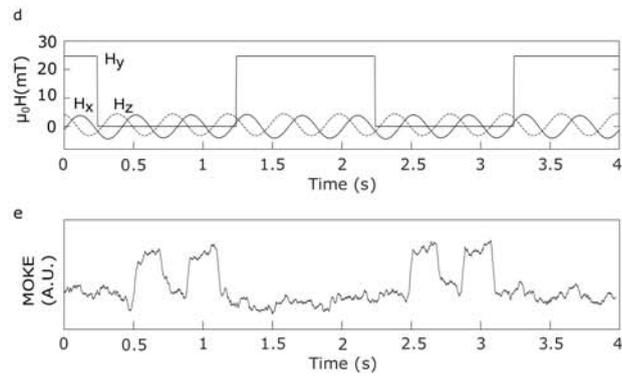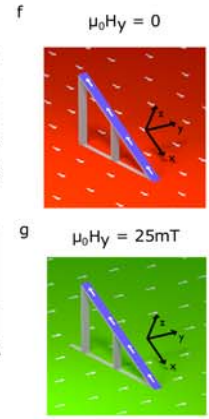